\documentclass[a4paper, twoside]{article}

\usepackage{xspace} % for \setype
\usepackage[usenames,dvipsnames,svgnames,table]{xcolor}
\usepackage{epsfig}
\usepackage{subfigure}
\usepackage{calc}
\usepackage{amssymb}
\usepackage{amstext}
\usepackage{multicol}
\usepackage{pslatex}
\usepackage{amsmath}
\usepackage{amsthm}
\usepackage{apalike}
\usepackage{hyperref}
\usepackage{breakurl}
\usepackage{SCITEPRESS}

\newcommand{\setype}[1]{{\tt #1}\xspace}
\newcommand{\p}{\phantom{0}}
\theoremstyle{definition}
\newtheorem{requirement}{R}

\newif\ifanonymous
\newif\ifnotanonymous
\newif\ifabridged
\newif\ifnotabridged

\def\isabridged{1}

\ifdefined\isabridged
\abridgedtrue
\fi

\ifabridged\notabridgedfalse
\else\notabridgedtrue
\fi

\ifdefined\isanonymous
\anonymoustrue
\fi

\ifanonymous\notanonymousfalse
\else\notanonymoustrue
\fi

\ifanonymous
\newcommand{\selint}{OTARY\xspace}
\else
\newcommand{\selint}{SELint\xspace}
\fi
\newcommand{\seal}{SEAL\xspace}
\newcommand{\selinux}{SELinux\xspace}
\newcommand{\seandroid}{SEAndroid\xspace}
\newcommand{\aosp}{AOSP\xspace}
\let\svthefootnote\thefootnote
\newcommand\blfootnote[1]{%
	\let\thefootnote\relax\footnotetext{#1}%
	\let\thefootnote\svthefootnote%
}

\begin{document}

\title{\selint\ifanonymous \thanks{Name anonymized for double-blind review} \fi: an \seandroid policy analysis tool}

% Authors
\ifnotanonymous
\author{
	\authorname{Elena Reshetova\sup{1}, Filippo Bonazzi\sup{2}, N. Asokan\sup{2,3}}
	\affiliation{\sup{1} Intel OTC, Helsinki, Finland}
	\affiliation{\sup{2} Aalto University, Helsinki, Finland}
	\affiliation{\sup{3} University of Helsinki, Helsinki, Finland}
	\email{elena.reshetova@intel.com, filippo.bonazzi@aalto.fi, asokan@acm.org}
}
\else
\author{
	\vspace*{6\baselineskip}
}
\fi

\keywords{Security, \seandroid, \selinux, Android, Access Control, policy analysis}

\abstract{\seandroid enforcement is now mandatory for Android devices.
	In order to provide the desired level of security for their products, Android OEMs need to be able to minimize their mistakes in writing \seandroid policies.
	However, existing \seandroid and \selinux tools are not very useful for this purpose.
	It has been shown that \seandroid policies found in commercially available devices by multiple manufacturers contain mistakes and redundancies.
	In this paper we present a new tool, \selint, which aims to help OEMs to produce better \seandroid policies.
	\selint is extensible and configurable to suit the needs of different OEMs.
	It is provided with a default configuration based on the \aosp \seandroid policy, but can be customized by OEMs.
}

%\onecolumn \maketitle \normalsize \vfill
\onecolumn \maketitle

%%%%%%%%%%%%%%%%%%%%%%%%%%%%%%%%%%%%%%%%%%%%%%%%%%%%%%%%%%%%%%%%%%%%%%%
% Start of the actual paper

\section{\uppercase{Introduction}}
\label{sec:introduction}

\blfootnote{Up-to-date version of this paper available at \href{https://arxiv.org/abs/1608.02339}{arxiv.org/abs/1608.02339}}
\noindent During the past decade Android OS has become one of the most common mobile operating systems.
However, at the same time we have seen a big increase in the number of malware and exploits available for it~\cite{zhou2012dissecting,smalley2013security}.
Many classical Android exploits, such as GingerBreak and Exploid, attempted to target system daemons that ran with elevated - often unlimited - privileges.
A successful compromise of such daemons results in the compromise of the whole Android OS, and the attacker would be able to obtain permanent root privileges on the device.
Initially Android relied only on its permission system, based on Linux Discretionary Access Control (DAC), to provide security boundaries.
However, after it became evident that DAC cannot protect from such exploits, a new Mandatory Access Control (MAC) mechanism has been introduced.
\seandroid~\cite{smalley2013security} is an Android port of the well-established \selinux MAC mechanism~\cite{smalley2001implementing} with some Android-specific additions.
Similarly to \selinux, \seandroid enforces a system-wide policy.
The default \seandroid policy was created from scratch and is maintained as part of the Android Open Source Project (\aosp)\footnote{\href{https://source.android.com}{source.android.com}}.
Starting from the 5.0 Lollipop release, Android requires every process to run inside a confined SEAndroid domain with a proper set of access control rules defined.
This has forced many Android Original Equipment Manufacturers (OEMs) to hastily define the set of access control domains and associated rules needed for their devices.
\ifanonymous A \else Our \fi recent study~\cite{reshetova2015characterizing} showed that all OEMs modify the default \seandroid policy provided by \aosp due to many customizations implemented in their Android devices.
The difficulty of writing well-designed \selinux policies together with high time-to-market pressure can possibly lead to the introduction of mistakes and major vulnerabilities.
The study classified common mistake patterns present in most OEM policies and concluded that new practical tools are needed in order to help OEMs avoid these mistakes.
In this paper we make the following contributions:

\begin{itemize}
	\item Design of \textbf{a new, extensible tool, \selint,} that aims to help Android OEMs to overcome common challenges when writing \seandroid policies (Section~\ref{sec:selint}). In contrast to existing \selinux and \seandroid tools (described in Section~\ref{sec:related-work}), it can be used by a person without a deep understanding of \seandroid, given the initial configuration by an expert.
	      The community can write new analysis modules for \selint in the form of \selint plugins.
	      This is especially important given that the \seandroid policy format changes with every release, and new notions and mechanisms are introduced by Google.
	\item \textbf{An initial configuration} for \selint, based on the \aosp \seandroid policy, which \textbf{was found to be useful} by the \seandroid community in our evaluation of \selint (Section~\ref{sec:userfeedback}).
	\item A full \textbf{implementation} of \selint that that \textbf{fits OEM policy development workflows}, providing \textbf{reasonable performance} and allowing \textbf{easy customization} by OEMs (Section~\ref{sec:evaluation}).
\end{itemize}

\section{Background}
\subsection{\selinux and \seandroid}
\label{sec:selinuxseandroid}

\textbf{\selinux}~\cite{smalley2001implementing} was the first mainline MAC mechanism available for Linux-based distributions.
Compared to other mainline MAC mechanisms present today in the Linux kernel, it is considered to be the most fine-grained and the most difficult to understand and manage due to the lack of a minimal policy (like in Smack~\cite{schaufler2008smack}) or a learning mode (like in AppArmor~\cite{bauer2006paranoid}).
Despite this, it is enabled by default in Red Hat Enterprise Linux (RHEL) and Fedora with pre-defined security policies.

The core part of \selinux is its Domain/Type Enforcement~\cite{badger1995practical} mechanism, which assigns a \setype{domain} to each subject, and a \setype{type} to each object in the system.
A subject running in \setype{domain} can only access an object belonging to \setype{type} if there is an \setype{allow} rule in the policy of the following form:

\begin{center}
	\setype{allow domain type : class permissions}
\end{center}

where \setype{class} represents the nature of an object such as file, socket or property, and \setype{permissions} represent the kinds of operations being permitted on this object, like \setype{read}, \setype{write}, \setype{bind} etc.

The \textbf{\seandroid}~\cite{smalley2013security} MAC mechanism is mostly based on \selinux code with some additions to support Android-specific mechanisms, such as the Binder Inter Process Communication (IPC) framework.
However, \seandroid's policy is fully written from scratch and is very different from \selinux's reference policy.
\aosp predefines a set of application domains, like \setype{system\_app}, \setype{platform\_app} and \setype{untrusted\_app}; applications are assigned to these domains based on the signature of the Android application package file (\texttt{.apk}).
Other services and processes are assigned to their respective domains based on filesystem labeling or direct domain declaration in the service definition in the \setype{init.rc} file.
One notable feature of the \seandroid policy is active usage of predefined M4 macros that make the policy more readable and compact.
For example, the \setype{global\_macros} file defines a number of M4 macros that denote sets of typical permissions needed for common classes, such as \setype{r\_file\_perms} or \setype{w\_dir\_perms}.
Another example is the \setype{te\_macros} file, that provides a number of M4 macros used to combine sets of rules commonly used together.

\subsection{\seandroid OEM Challenges}
\label{sec:oem-challenges}

The \seandroid reference policy only covers default \aosp services and applications.
Therefore, highly customized OEM Android devices require extensive policy additions.

\ifanonymous The \else Our \fi already mentioned study of different OEM \seandroid policies for Android 5.0 Lollipop~\cite{reshetova2015characterizing} showed that most OEMs made a significant number of additions to the default \aosp reference policy.
The biggest changes are the additions of new types and domains, as well as new \setype{allow} rules.
The study also identified a number of common patterns that most OEM policies seem to follow:

\begin{itemize}
	\item \textbf{Overuse of default types}. \seandroid declares a set of default types that are assigned to different objects unless a dedicated type is defined in the policy.
	      Most OEMs leave many such types in their policies, which indicates a use of automatic policy creation tools such as \setype{audit2allow}~\cite{selinuxuserspace}.
	\item \textbf{Overuse of predefined domains}. OEMs do not typically define dedicated domains for their system applications, but tend to assign these applications to predefined \setype{platform\_app} or \setype{system\_app} domains.
	      This creates over-permissive application domains and violates the principle of least privilege.
	\item \textbf{Forgotten or seemingly useless rules}. OEM policies have many rules that seem to have no effect.
	      This might be due to an automatic rule generation or a failure to clean up unnecessary rules that were no longer required.
	\item \textbf{Potentially dangerous rules}. A number of potentially dangerous rules can be seem in some OEM policies, including granting additional permissions to \setype{untrusted\_app} domain.
	      This might be due to lack of time to adjust their service or application implementation to minimize security risks or due to inability to identify some rules as being dangerous.
\end{itemize}

\section{\uppercase{Related Work}}
\label{sec:related-work}

\noindent Since \selinux existed on its own long before \seandroid, most of the available tools are designed to handle and analyze \selinux policies.
They can be used for \seandroid but they don't take specific aspects of \seandroid policies into account.
This makes it challenging for OEMs to use existing tools to detect the problems outlined in Section~\ref{sec:oem-challenges}.
For example, in order to determine if the policy contains potentially dangerous rules, it is very important to understand the semantics of \seandroid types and policy structure - an ability which all existing \selinux tools lack.
Moreover, even the small group of \seandroid tools described in Section~\ref{sec:seandoid-tools} does not address the challenges described in Section~\ref{sec:oem-challenges}.

\subsection{\selinux Tools}

SETools~\cite{setools4} is the official collection of tools for handling \selinux policies in text and binary format.
Some of its tools, like \texttt{apol}, are suitable for formal policy analysis, for example for flow-control analysis.
Others allow policy queries and policy parsing and as such it can be used on both \selinux and \seandroid.
An important part of SETools is a policy representation library which is used in both \seal and \selint.

Formal methods have been applied to \selinux policy analysis.
Gokyo~\cite{jaeger2003analyzing} is a tool designed to find and resolve conflicting policy specifications.
Guttman \textit{et al.}~\cite{guttman2005verifying} applies information flow analysis to \selinux policies.
The HRU security model~\cite{Harrison:1976:POS:360303.360333} has been used to analyze \selinux policies~\cite{amthor2011model}.
Hurd \textit{et al.}~\cite{hurd2009policy} applied Domain Specific Languages (DSL)~\cite{fowler2010domain} in order to develop and verify the \selinux policy.
The resulting tool, \texttt{shrimp}, can be used to analyze and find errors in the \selinux Reference Policy.
Information visualization techniques have been applied to \selinux policy analysis in~\cite{clemente2012sptrack}, also in combination with clustering of policy elements~\cite{marouf2011segrapher}.
These analysis methods are largely academic, and no practical tools based on them are used by the \selinux community.

Polgen~\cite{sniffen2006guided} is a tool for semi-automated \selinux policy generation based on system call tracing.
Unfortunately it appears to be no longer in active development.
\selinux also provides a set of userspace tools~\cite{selinuxuserspace} that can be used on both \selinux and \seandroid.
One of these tools, \texttt{audit2allow}, is widely used by Android OEMs to automatically generate and expand \seandroid policies.
The tool works by converting denial audit messages into rules based on a given binary policy.
These rules, however, are not necessarily correct, complete or secure, since they entirely depend on code paths taken during execution and require a good understanding of the software components involved, as well as on the correct labeling of subjects and objects in the system.
Furthermore, automatically-generated rules fail to use high-level \seandroid policy features such as attributes and M4 macros: this results in comparatively less readable policies.

\subsection{\seandroid Tools}
\label{sec:seandoid-tools}

\ifanonymous The \else Our \fi aforementioned study~\cite{reshetova2015characterizing} presented \seal, an \seandroid live device analysis tool.
\seal works with a real or emulated Android device over the Android Debug Bridge (ADB); it can perform different queries that take into account not only the binary \seandroid policy loaded on the device, but also the actual device state, i.e. running processes and filesystem objects.
The EASEAndroid policy refinement method is based on audit log analysis with machine learning~\cite{wang15easeandroid}.
This approach is completely different from what we propose, since it relies on significant volumes of data to classify rules.
Unfortunately, it is very hard to obtain this volume of data, since it would require collecting log files from millions of Android devices with possible privacy implications.
The most recent \seandroid policy analysis and refinement tool is \seandroid Policy Knowledge Engine (SPOKE) ~\cite{wang2016automatic}.
It automatically extracts domain knowledge about the Android system from application functional tests, and applies this knowledge to analyze and highlight potentially over-permissive policy rules.
SPOKE can be used to identify new heuristics that can be implemented as new \selint plugins.
The downside of SPOKE is reliance on application functional tests, which are often incomplete, and the fact that it cannot be easily integrated into the standard development workflow.

\section{\uppercase{\selint}}
\label{sec:selint}

\subsection{Requirements}
\label{sec:requirements}

We identify the following generic requirements that a tool like \selint must fulfill.

\begin{requirement}
	\label{req:source}
	\textbf{Source policy-based.} The existing tool landscape presented in Section~\ref{sec:related-work} does not feature any tool able to perform semantic analysis on source \seandroid policies.
	Since Android OEMs work on source \seandroid policies as part of their Android trees, the tool needs to work with source \seandroid policies.
\end{requirement}
\begin{requirement}
	\label{req:config}
	\textbf{Configurable by experts, usable by all.} Existing tools require extensive domain knowledge to be used.
	Since building such a knowledge takes considerable time, it might be challenging for OEMs to have all of their development team trained appropriately.
	We intended for our tool to fit into an Android OEM policy development workflow, where many developers, overseen by one or a few experienced \seandroid analysts, contribute small changes to the policy.
	Therefore, it must be possible for an experienced analyst to configure the tool ahead of time, and provide a ready-to-run tool to regular developers, who can simply run the tool on their policy modifications and verify that no issues are highlighted.
\end{requirement}
\begin{requirement}
	\label{req:performance}
	\textbf{Reasonable performance.} Since we are targeting inclusion into an Android OEM workflow, the tool must have reasonable time and memory performance; this is necessary for the tool to be used as part of the build toolchain, or even more appropriately when committing changes using the OEM's version control software (VCS).
\end{requirement}
\begin{requirement}
	\label{req:extend}
	\textbf{Easy to configure and extend.} Finally, targeting the wide community of Android OEMs makes it impossible to know in advance all possible use cases and requirements, present and future.
	It is our objective to allow analysts to implement their own analysis functionality and embed their domain knowledge into the tool.
	For this reason, the tool must be easily configurable and extensible by the community.
\end{requirement}

\subsection{General Architecture and Implementation}
To meet Requirement~\ref{req:extend} stated in section~\ref{sec:requirements}, we designed \selint following a plugin architecture.
The goal of such an architecture is to support custom third-party analysis plugins that any community member can create.
The core part of \selint is responsible for processing the source \seandroid policy.
The \emph{\selint core} takes care of handling user input, such as command line options and configuration files.
After the source policy has been parsed, its representation is given to the \emph{\selint plugins} which perform the actual analysis.
We have developed an initial set of plugins, which provide generally useful functionality; interested Android OEMs can develop more plugins to implement their own analysis requirements.

\begin{figure}[h]
	\ifanonymous
	\includegraphics[width=\columnwidth]{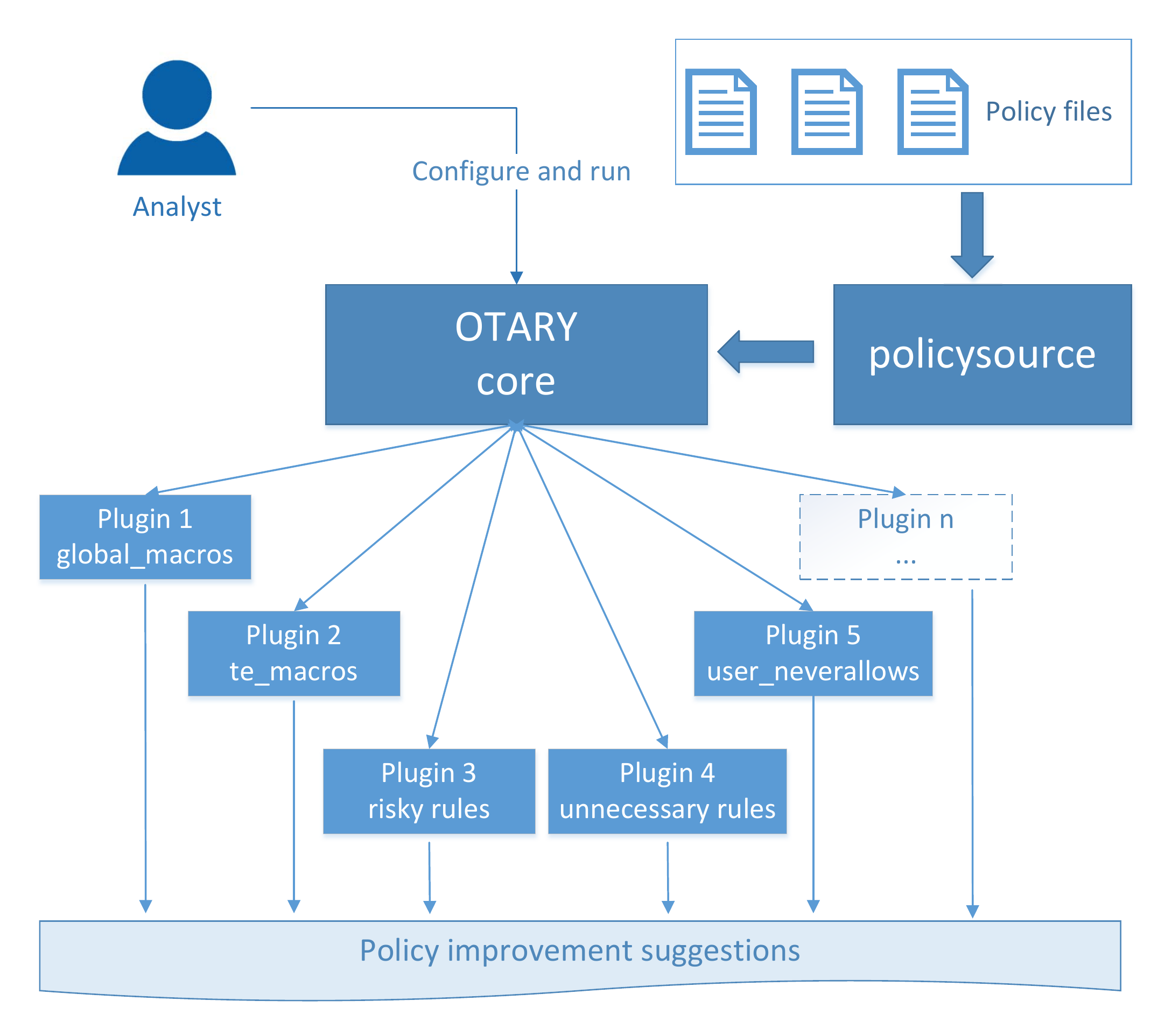}
	\else
	\includegraphics[width=\columnwidth]{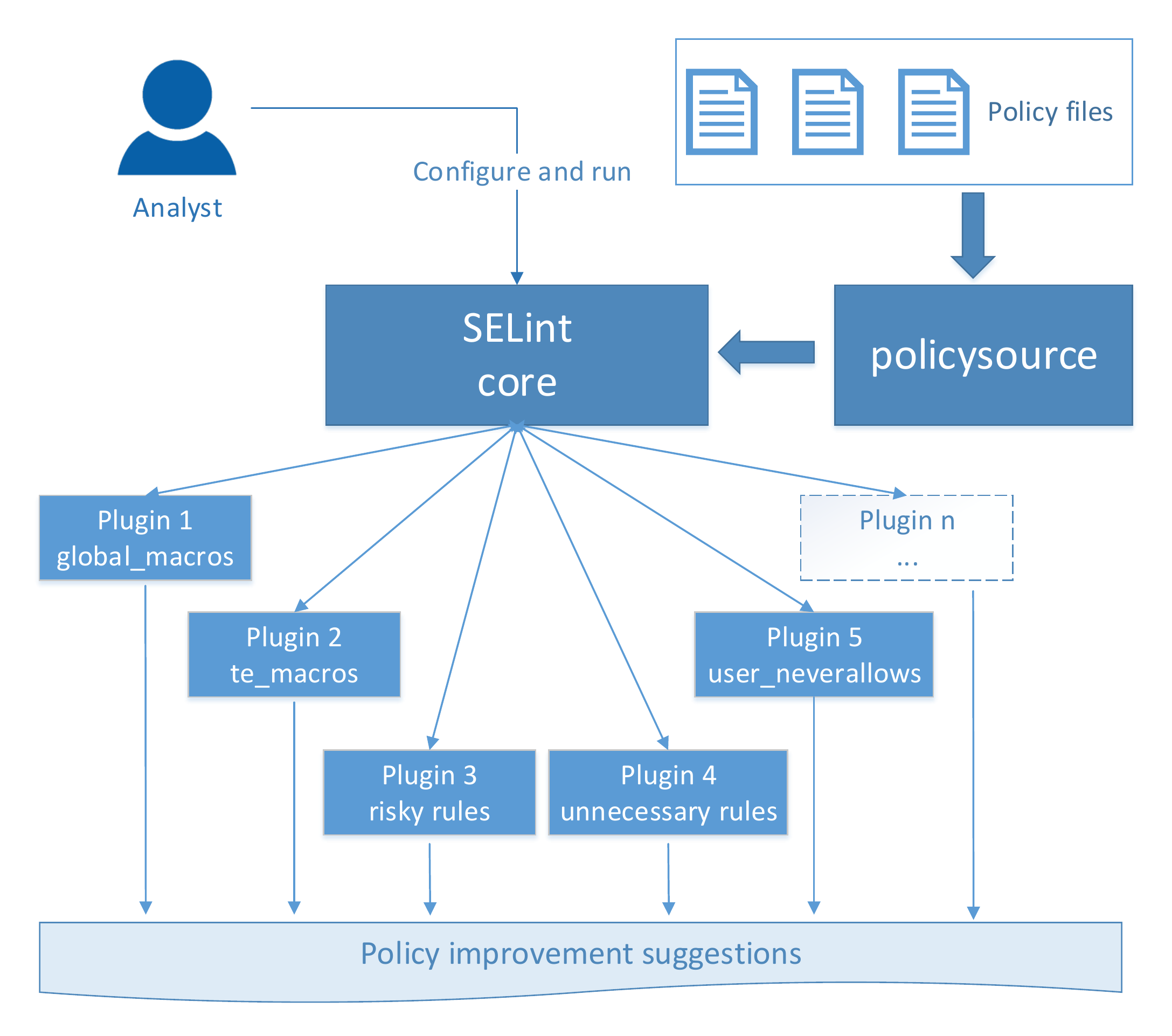}
	\fi
	\caption{The architecture of \selint}
	\label{fig:selint-archi}
\end{figure}

The overall architecture is shown in Figure~\ref{fig:selint-archi}; the existing plugins are individually described in the following sections.
The implementation of \selint and the existing plugins are released under the Apache License 2.0, which allows the community to freely use and modify the software.
The \texttt{policysource} library is released under the GNU Lesser General Public License v2.1.

The \selint executable and all plugins have an associated configuration file.
This allows policy experts to adapt each plugin to the semantics of their own policies, for example to define OEM-specific policy types.
This way, \selint can be run with different preset ``\emph{profiles}'' specifying  different options, policy configurations and requested analysis functionalities.
The following sections describe each existing \selint plugin in detail.

\subsection{Plugin 1: Simple Macros}
\textbf{Goal}
\begin{figure}[h]
	\resizebox{\columnwidth}{!}{%
		\begin{tabular}{l}
			\texttt{r\_file\_perms $\rightarrow$ \{ getattr open read ioctl lock \}} \\
		\end{tabular}
	}
	\caption{A \texttt{global\_macros} definition and expansion}
	\label{fig:global-macro-example}
\end{figure}
As mentioned in Section~\ref{sec:selinuxseandroid}, using M4 macros where applicable is a non-functional requirement of \seandroid policy development: while not affecting policy behavior, their use makes for a more compact and readable policy.
The first type of M4 macros extensively used in \seandroid policies is a simple text replacement macro, without arguments, that is used to represent sets of related permissions.
Such macros are defined in the \texttt{global\_macros} file in the \seandroid policy source files.
An example of such macro is shown in Figure~\ref{fig:global-macro-example}.
The Simple macro plugin scans the policy for rules granting sets of individual permissions which could be represented in a more compact way by using an existing \texttt{global\_macros} macro; it then suggests replacing the individual permissions with an usage of said macro.

\noindent\textbf{Implementation}
The plugin looks for rules which specify individual permissions whose combination is equivalent to the expansion of a \texttt{global\_macros} macro.
It then suggests rewriting said rules, replacing the individual permissions with the unexpanded macro.
An example is shown in Figure~\ref{fig:global-macros-plugin-example}.
For this particular case, the plugin suggest replacing a set of permissions \texttt{\{gettattr open read search ioctl\}} with a macro \texttt{r\_dir\_perms}.
Permissions not contained in the macro (in this case \texttt{create}) are still specified individually in the final rule.
The plugin can suggest both full matches (for rules that grant 100\% of the permissions contained in a macro) and partial matches above a threshold (for rules that grant at least X\% of the permissions contained in a macro).
This threshold is a user-defined parameter, specified in the plugin configuration file; we assigned it a default value of 0.8 (80\%).

\begin{figure}[h]
	\resizebox{\columnwidth}{!}{%
		\begin{tabular}{l}
			\textbf{Rule:}                                                                                       \\
			\texttt{allow logd rootfs:dir}                                                                       \\
			\texttt{    \{ \textcolor{gray}{getattr} create \textcolor{gray}{open read search ioctl} \};}        \\
			                                                                                                     \\
			\textbf{Macro:}                                                                                      \\
			\texttt{r\_dir\_perms} $\rightarrow$ \texttt{\{ \textcolor{gray}{open getattr read search ioctl} \}} \\
			                                                                                                     \\
			\textbf{Suggestion:}                                                                                 \\
			\texttt{allow logd rootfs:dir} \texttt{\{ r\_dir\_perms create \};}                                  \\
		\end{tabular}
	}
	\caption{An example usage of the Simple macro plugin}
	\label{fig:global-macros-plugin-example}
\end{figure}

\noindent\textbf{Limitations}
The plugin only deals with simple, static macros without arguments.
Dynamic macros such as those defined in the \texttt{te\_macros} file are handled by the dedicated plugin described in the next section.

\subsection{Plugin 2: Parametrized Macros}
\label{sec:te-macro}

\textbf{Goal}
\begin{figure}[h]
	\centering
	\small
	\texttt{
		\begin{tabular}{l}
			\p\p\p\p\p\p\p`file\_type\_trans(\$1, \$2, \$3)'                             \\
			\p\p\p\p\p\p\p\p\p\p\p\p\p\p\p\p\p\p\p\p$\downarrow$                         \\
			`allow \$1 \$2:dir ra\_dir\_perms;                                           \\
			\p allow \$1 \$3:dir create\_dir\_perms;                                     \\
			\p allow \$1 \$3:notdevfile\_class\_set                                      \\
			\p\p\p\p\p\p\p\p\p\p\p\p\p\p\p\p\p\p\p\p\p\p\p\p\p\p\p create\_file\_perms;' \\
		\end{tabular}
	}
	\caption{A \texttt{te\_macros} macro definition and expansion with arguments.}
	\label{fig:te-macro-example}
\end{figure}
Another commonly used set of M4 macros includes more complex, dynamic M4 macros with multiple arguments.
Such macros are mainly used to group rules which are commonly used together; their expansion can in turn contain other macros.
In \seandroid policies, such macros are defined in the \texttt{te\_macros} file.
An example is shown in Figure~\ref{fig:te-macro-example}.
Similarly to the previous, this plugin detects existing macro definitions, and suggests new usages.

\noindent\textbf{Implementation}
The plugin looks for sets of individually specified rules whose combination is equivalent to the expansion of a \texttt{te\_macros} macro with some set of arguments.
It then suggests substituting said rules with a usage of the unexpanded macro with the proper arguments.
An example is shown in Figure~\ref{fig:te-macros-plugin-example}: the plugin finds the existing macro which expands into the given set of rules - in this case, \texttt{unix\_socket\_connect}.
It then extracts the arguments from the rules: \$1 is ``a'', \$2 is ``b'' and \$3 is ``c''.
The result is a suggestion for substituting the two rules with the macro usage.
The plugin can suggest both full matches (for sets of rules that match 100\% of the rules contained in a macro expansion) and partial matches above a user-defined threshold.
Its default value is 0.8 (80\%).

\begin{figure}[h]
	\centering
	\resizebox{\columnwidth}{!}{%
		\begin{tabular}{l}
			\textbf{Rules:}                                        \\
			\texttt{allow a b\_socket:sock\_file write;}           \\
			\texttt{allow a c:unix\_stream\_socket connectto;}     \\
			                                                       \\
			\textbf{Macro:}                                        \\
			\texttt{unix\_socket\_connect(\$1, \$2, \$3)}          \\
			\p\p\p\p\p\p\p\p\p\p\p\p\p\p\p$\downarrow$             \\
			\texttt{allow \$1 \$2\_socket:sock\_file write;}       \\
			\texttt{allow \$1 \$3:unix\_stream\_socket connectto;} \\
			                                                       \\
			\textbf{Suggestion:}                                   \\
			\texttt{unix\_socket\_connect(a, b, c)}                \\
		\end{tabular}
	}
	\caption{An example usage of the Parametrized macro plugin.}
	\label{fig:te-macros-plugin-example}
\end{figure}

\noindent\textbf{Limitations}
The problem of detecting sets of rules that match possible macro expansions can be transformed into a variant of the knapsack problem~\cite{kellerer2004knapsack}, namely a multidimensional knapsack problem.
In our case, the knapsack capacity is the number of arguments a macro can have, and the knapsack items are the possible values of these arguments; the knapsack is multidimensional because filling an argument does not affect the available capacity for the others.
Instead of finding the single most profitable combination of argument-values, our objective is to find all the combinations of argument-values which, used as arguments in as many macro expansions, produce sets of rules entirely or partially (above a threshold) contained in the policy.
The problem can be formalized as:

\noindent \textbf{For each macro $m$, find all combinations of values for arguments \$1, \$2 and \$3 such that $y$ is above an user-given threshold $t$.}
$y$ is computed as: $y = score(m(i, j, k))$, subject to $i \in N_i, j \in N_{j}|_i, k \in N_{k}|_{ij}$, where $N_i$ is the set of possible values of \$1, $N_{j}|_i$ is the set of possible values of \$2 given $i$ as \$1, and $N_{k}|_{ij}$ is the set of possible values of \$3 given $i$ as \$1 and $j$ as \$2.
$score(m(i, j, k))$ is the score of the macro expanded with the arguments $i$, $j$ and $k$: the score of a macro expansion is given by the number of its rules actually found in the policy divided by its overall number of rules.

The multidimensional knapsack optimization problem is known to be NP-hard~\cite{magazine1984note}, and it has various approximate solutions~\cite{chu1998genetic,hanafi1998efficient}.
In our case, the problem quantities are the number of arguments a macro can have (existing macros have 1-3), the number of rules a macro expansion can produce (existing macros have 1-7), and the number of values a macro argument can have (in principle infinite, in practice dependent on the policy, usually in the thousands).
In practice, the number of rules (\#2) tends to increase linearly with the number of arguments (\#1).
This is due to the fact that macros with more arguments can define more complex behavior, which tends to be described in more rules.

As a first implementation, we realized a simple solution based on exploration of the solution space: we try to aggregate all the policy rules into sets corresponding to macro expansions.
The problem quantities described above result in a significant time expenditure required to explore the whole solution space: therefore, as we discuss in Section~\ref{sec:performance}, this plugin takes considerably more time than all others.

\subsection{Plugin 3: Risky Rules}
\label{sec:risky-rules}

Experts analyzing OEM modifications to \seandroid policies often use certain heuristics.
The analysis usually starts from the list of \aosp \seandroid domains and types that are more likely to cause potential vulnerabilities in OEM policies.
The most common are:

\begin{itemize}
	\item \textbf{Untrusted domains.} Some domains are intended to run potentially malicious code, such as \texttt{untrusted\_app}, and therefore their privileges are designed to be minimal.
	      Any additional \setype{allow} rules created by OEMs for such domains are suspicious and need to be analyzed.
	\item \textbf{Trusted Computing Base (TCB) domains and types.} The \aosp policy has several core domains and types, which form its TCB.
	      The processes that run in these domains are provided by \aosp, and so are the minimal required policy rules.
	      Sometimes, OEMs have to create additional rules for some of these domains: however, since doing so increases the chance of compromising the TCB, such rules need thoughtful inspection.
	\item \textbf{Security-related domains and types.} Special attention must be paid to \aosp domains and types directly related to system security, such as the \texttt{tee} domain or the \texttt{proc\_security} type.
	      Mistakes in additional \setype{allow} rules for these domains and types can lead to a direct loss of system security.
\end{itemize}

An analyst usually checks an OEM policy for additional rules where the above domains or types are present, and then manually inspects each rule analysing its domain, type and permissions to determine if the rule is actually risky.
This process is tedious, and most of the time is spent just finding the rules which need special attention.
To help analysts find these rules quickly, we developed the \texttt{risky\_rules} plugin, which processes each rule and assigns it a score based on one of two criteria.

The first scoring criterion is based on \textit{risk}.
We define the \textbf{risk level} for rule components as the level of potential damage to the system caused by misuse of the component: security-sensitive components will have high risk scores, while generic components will have lower risk scores.
Untrusted domains will have a high \textit{risk} score as well, because we want to select any additional rules over such domains for manual inspection.
Component risk level in turn determines the risk level of a rule, which is obtained by combining the risk levels of its components.
The risk level of a rule is then defined as the level of potential damage to the system allowed by the rule.
The risk score helps analysts to quickly obtain a prioritized list of policy rules which need manual inspection; this is especially useful when analysts have strict time constraints, and only have time to examine a limited number of rules.
The \textit{risk} scoring system is described in Section~\ref{sec:risky-rules-risk-scoring-system}.

The second scoring criterion is based on \textit{trust}.
We define the \textbf{trust level} for rule components as a measure of closeness to the core of the system: key system components will have a high \textit{trust} level, while user applications will have a low \textit{trust} level.
This in turn allows us to detect rules which cross trust boundaries, \textit{e.g.} comprising a \textit{high} component and a \textit{low} component or vice versa.
This scoring system is useful for an analyst as well, because it can quickly identify additional OEM rules which breach trust boundaries and select them for manual inspection.
The \textit{trust} scoring system is described in Section~\ref{sec:risky-rules-trust-scoring-system}.

The desired scoring system can be specified in the plugin configuration file.
We have provided an initial \texttt{risky\_rules} plugin configuration based on our knowledge and experience with the \aosp policy.
While our classification might be considered subjective, feedback discussed in Section~\ref{sec:userfeedback} indicates that \seandroid policy writers agree with our approach.

\subsubsection{Measuring Risk}
\label{sec:risky-rules-risk-scoring-system}
\textbf{Goal}
As mentioned above, rules in a policy can have different risk levels, depending on the types they deal with and the permissions they grant.
The \textit{risk} scoring system of the \texttt{risky\_rules} plugin assigns a score to every rule in the policy, prioritizing potentially riskier rules by assigning them higher scores.

\noindent\textbf{Implementation}
The \textit{risk} scoring system computes the overall score for a rule by evaluating its domain, type, and permissions or capabilities.
The plugin configuration file defines partial \textit{risk} scores for various rule elements.
Relevant \aosp domains and types are grouped by risk level into ``\textit{bins}'', which are assigned a partial \textit{risk} score with a maximum of 30.
When computing the score for a rule, the partial scores of its domain and type are added.
We treat domains and types equally, because both the running process and data of a program might be equally important in evaluating how risky a rule is.
For example, a process running in a security sensitive domain (\textit{e.g.} \texttt{keystore}) should not accept any command from other processes running in unauthorized domains, because they might induce malicious changes in its execution flow.
Similarly, other unauthorized processes should not be able to modify the configuration data of a security sensitive process (\textit{e.g.} data labeled as \texttt{keystore\_data}), for similar reasons.
The initial set of bins and their default scores are depicted in Table~\ref{tab:risky-rules-risk-bins}.

\begin{table}[h]
	\centering
	\caption{\texttt{risky\_rules} plugin default bins and partial \textit{risk} scores.}
	\resizebox{\columnwidth}{!}{
		\begin{tabular}{|c | c | c |} \hline
			\textbf{Bin name}            & \textbf{Example types}      & \textbf{Risk} \\ \hline
			\texttt{user\_app}           & \texttt{untrusted\_app}     & 30            \\ \hline
			\texttt{security\_sensitive} & \texttt{tee, keystore},     & 30            \\
			                             & \texttt{security\_file}     &               \\ \hline
			\texttt{core\_domains}       & \texttt{vold, netd, rild}   & 15            \\ \hline
			\texttt{default\_types}      & \texttt{device, unlabeled}, & 30            \\
			                             & \texttt{system\_file}       &               \\ \hline
			\texttt{sensitive}           & \texttt{graphic\_device}    & 20            \\ \hline
		\end{tabular}
	}
	\label{tab:risky-rules-risk-bins}
\end{table}

The \texttt{user\_app}, \texttt{core\_domains} and \texttt{security\_sensitive} bins match groups defined earlier in this section.
\texttt{user\_app} and \texttt{security\_sensitive} have the maximum score of 30, while the score for \texttt{core\_domains} is 15 due to less overall risk to the system.
The \texttt{default\_types} bin has a maximum score of 30, because it contains types that should not normally be used by OEMs and therefore likely indicate a mistake in a rule.

When computing the overall \textit{risk} score for a rule, in addition to evaluating a rule's domain and type elements, the \textit{risk} scoring system must also take its permissions and capabilities into account.
In \seandroid, permissions are meaningless in isolation, and only meaningful to determine risk when combined with the domain to which they are granted and the type over which they are granted: for this reason, we combine these when computing the \textit{risk} score for a rule.
We do this by assigning permissions a multiplicative coefficient instead of an additive partial score; the sum of domain and type score for a rule is multiplied by this coefficient.
Commonly used permissions are categorized by level of risk into three groups, \setype{perms\_high}, \setype{perms\_med} and \setype{perms\_low}: each group is assigned a coefficient based on the sensitivity of its permissions, with a maximum of 1.
The sum of domain and type score is multiplied by the coefficient of the highest set which contains permissions granted by the rule; this is done because we are interested in determining the upper bound of risk for a rule.
Table~\ref{tab:risky-rules-perms} shows the groups, permissions and default values of coefficients.

\begin{table}[h]
	\centering
	\caption{\texttt{risky\_rules} plugin default permission sets and coefficients.}
	\resizebox{\columnwidth}{!}{
		\begin{tabular}{|c | c | c|} \hline
			\textbf{Set name}    & \textbf{Example permissions}   & \textbf{Coefficient} \\ \hline
			\texttt{perms\_high} & \texttt{ioctl, write, execute} & 1                    \\ \hline
			\texttt{perms\_med}  & \texttt{read, use, fork}       & 0.9                  \\ \hline
			\texttt{perms\_low}  & \texttt{search, getattr, lock} & 0.5                  \\ \hline
		\end{tabular}
	}
	\label{tab:risky-rules-perms}
\end{table}

Capabilities are treated differently from permissions.
In \seandroid, capabilities are granted by a domain to itself, and - unlike permissions - are meaningful on their own: they have the same effect on the system regardless of the domain they are granted to.
For example, the following rule grants the \texttt{vold} daemon the \texttt{CAP\_CHROOT} capability, which allows it to perform the \texttt{chroot} system call:

\begin{center}
	\texttt{allow vold self:capability sys\_chroot;}
\end{center}

We do not divide capabilities into separate groups: this is due to the fact that, in Linux, capabilities are commonly believed to be very hard to categorize as more or less dangerous, because of the consequences they can have on the system\footnote{\href{https://forums.grsecurity.net/viewtopic.php?f=7&t=2522}{forums.grsecurity.net/viewtopic.php?f=7\&t=2522}}.
Since in \seandroid capabilities are granted by a domain to itself, the target type in such a rule does not convey any additional information: therefore, we use a special scoring formula for rules granting capabilities.
Capabilities are handled as types, and any capability is assigned the maximum score for a type (30): this score is added to the domain score to obtain the rule score.

\begin{figure}[h]
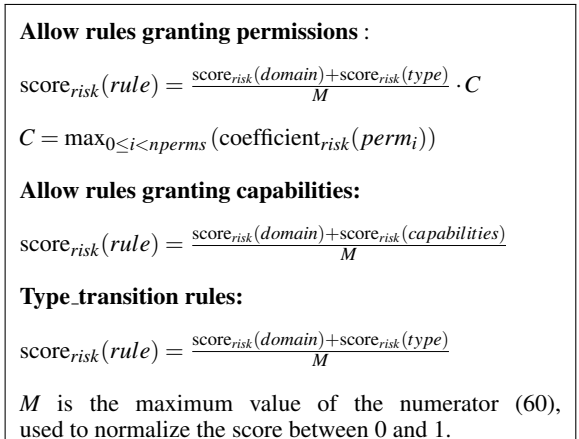

	\small
	\fbox{
		\begin{minipage}{0.95\columnwidth}
			\begin{math}
				\\
				\textbf{Allow rules granting permissions}:\\
				\\
				\text{score}_{risk}(rule) = {{\text{score}_{risk}(domain) + \text{score}_{risk}(type)} \over M} \cdot C\\
				\\
				C = \max_{0 \leq i < nperms}{(\text{coefficient}_{risk}(perm_i))}
				\\
				\\
				\textbf{Allow rules granting capabilities:}\\
				\\
				\text{score}_{risk}(rule) = {{\text{score}_{risk}(domain) + \text{score}_{risk}(capabilities)} \over M}
				\\
				\\
				\textbf{Type\_transition rules:}\\
				\\
				\text{score}_{risk}(rule) = {{\text{score}_{risk}(domain) + \text{score}_{risk}(type)} \over M}
				\\
			\end{math}
			\\
			$M$ is the maximum value of the numerator (60), used to normalize the score between 0 and 1.
		\end{minipage}
	}
	\caption{The \textit{risk} scoring formulas for the \texttt{risky\_rules} plugin.}
	\label{fig:risk-scoring-formulas}
\end{figure}

The \textit{risk} scoring system scores rules by their potential level of risk between 0 and 1, with maximum risk given a score of 1.
As discussed above, risk scores are assigned to rules depending on the type of rule: the precise formulas are presented in Figure~\ref{fig:risk-scoring-formulas}.

An example is shown in Figure~\ref{fig:risky-rules-plugin-risk-example}.
The first rule contains \setype{untrusted\_app} and \setype{security\_file}, which are both high-risk types (\setype{user\_app} and \setype{security\_sensitive} respectively); however, the rule only grants the \setype{getattr} and \setype{search} permissions, which are two low-risk permissions.
Thus, the rule has a medium \textit{risk} score that in this case equals to 0.5.
The second rule contains \setype{untrusted\_app} and \setype{system\_file}, which are both high-risk types (\setype{user\_app} and \setype{default\_types} respectively); furthermore, the rule grants the \setype{execute} permission, which is a high-risk permission.
Thus, the rule has a high \textit{risk} score that in this case equals to 1.

\begin{figure}[h]
	\small
	\centering
	\begin{verbatim}
0.50: .../domain.te:154: allow untrusted_app
        security_file:dir { getattr search };
1.00: .../domain.te:104:
    allow untrusted_app system_file:file execute;
	\end{verbatim}
	\caption{An example of the \texttt{risky\_rules} plugin with the \textit{risk} scoring system.}
	\label{fig:risky-rules-plugin-risk-example}
\end{figure}

\subsubsection{Measuring Trust}
\label{sec:risky-rules-trust-scoring-system}
\textbf{Goal}
Rules in a policy can contain domains and types with different \textit{trust} levels.
Analysts usually inspect a policy by manually looking for rules which cross \textit{trust} boundaries and making sure they are justified: this process is time-consuming and can be error prone.
The \textit{trust} scoring system of the \texttt{risky\_rules} plugin automates this search: it assigns a score to every rule in the policy, prioritizing rules which cross \textit{trust} boundaries by assigning them higher scores.

\noindent\textbf{Implementation}
The \textit{trust} scoring system combines the partial scores of domain and type in a rule to assign it an overall score.
The plugin configuration file defines partial \textit{trust} scores for various rule elements.
\aosp domains and types are grouped into ``\textit{bins}'', which are assigned a \textit{trust} score with a maximum of 30.
When computing the score for a rule, the partial scores of its domain and type are added.
The initial bins with their default scores are depicted in Table~\ref{tab:risky-rules-trust-bins}.

\begin{table}[h]
	\centering
	\caption{\texttt{risky\_rules} plugin default bins and partial \textit{trust} scores.}
	\resizebox{\columnwidth}{!}{
		\begin{tabular}{| c | c | c |} \hline
			\textbf{Bin name}            & \textbf{Example types}      & \textbf{Trust} \\ \hline
			\texttt{user\_app}           & \texttt{untrusted\_app}     & 0              \\ \hline
			\texttt{security\_sensitive} & \texttt{tee, keystore},     & 30             \\
			                             & \texttt{security\_file}     &                \\ \hline
			\texttt{core\_domains}       & \texttt{vold, netd, rild}   & 20             \\ \hline
			\texttt{default\_types}      & \texttt{device, unlabeled}, & 5              \\
			                             & \texttt{system\_file}       &                \\ \hline
			\texttt{sensitive}           & \texttt{graphic\_device}    & 10             \\ \hline
		\end{tabular}
	}
	\label{tab:risky-rules-trust-bins}
\end{table}

For example, the \setype{user\_app} bin contains types assigned to generic user applications, such as \setype{untrusted\_app}; since user applications are not trusted, the \textit{trust} score for this bin is minimum (0).
The \setype{security\_sensitive} bin contains types assigned to data or components that have direct security impact, such as \setype{tee}, \setype{keystore}, \setype{proc\_security} \textit{etc}.
These components and their data are also highly trusted, since they form the TCB of the system, and therefore their \textit{trust} score is maximum (30).
The \textit{trust} scoring system scores rules by the level of trust of their domain and type, regardless of the type of rule.
Permissions and capabilities are ignored when computing the \textit{trust} score for a rule.
The level of trust can be \textit{high} or \textit{low}, giving place to 4 different scoring criteria: \textit{trust\_hl}, where the rule features a \textit{high} domain and a \textit{low} type, \textit{trust\_lh}, where the domain is \textit{low} and the type is \textit{high}, \textit{trust\_hh}, where both are \textit{high}, and \textit{trust\_ll}, where both are \textit{low}.
The various \textit{trust} criteria score rules between 0 and 1, where a score of 1 indicates that a rule is closest to the specified criterion.
A high rule score is obtained naturally when looking for \textit{high} components: to obtain a high rule score when looking for \textit{low} components, the component partial score is subtracted from the maximum partial score before normalizing.
Trust scores are assigned to rules using the formulas presented in Figure~\ref{fig:trust-scoring-formulas}.

\begin{figure}[h]
	\small
	\fbox{
		\begin{minipage}{0.95\columnwidth}
			\begin{math}
				\\
				\textbf{Trust\_ll:}\\
				\text{score}_{trust}(rule) = {{({M\over2} - \text{score}_{trust}(domain)) + ({M\over2} - \text{score}_{trust}(type))} \over M}
				\\
				\\
				\textbf{Trust\_lh:}\\
				\text{score}_{trust}(rule) = {{({M\over2} - \text{score}_{trust}(domain)) + (\text{score}_{trust}(type))} \over M}
				\\
				\\
				\textbf{Trust\_hl:}\\
				\text{score}_{trust}(rule) = {{(\text{score}_{trust}(domain)) + ({M\over2} - \text{score}_{trust}(type))} \over M}
				\\
				\\
				\textbf{Trust\_hh:}\\
				\text{score}_{trust}(rule) = {{\text{score}_{trust}(domain) + \text{score}_{trust}(type)} \over M}
				\\
			\end{math}
			\\
			$M$ is the maximum value of the numerator (60), used to normalize the score between 0 and 1.
		\end{minipage}
	}
	\caption{The \textit{trust} scoring formulas for the \texttt{risky\_rules} plugin.}
	\label{fig:trust-scoring-formulas}
\end{figure}

An example of one of the \textit{trust} scoring systems (\textit{trust\_lh}) is shown in Figure~\ref{fig:risky-rules-plugin-trust-example}.
The first rule contains \setype{untrusted\_app}, which is a low-trust domain, and \setype{system\_file}, which is a low-trust domain.
The scoring criterion assigns the maximum score to rules with a \textit{low} domain and a \textit{high} type: therefore, the rule has a medium \textit{trust\_lh} score, which in this case is 0.58.
The second rule contains \setype{untrusted\_app}, which is a low-trust domain, and \setype{security\_file}, which is a high-trust type.
According to the selected scoring criterion, the rule has the maximum \textit{trust\_lh} score of 1.

\begin{figure}[h]
	\centering
	\small
	\begin{verbatim}
0.58: .../domain.te:104:
    allow untrusted_app system_file:file execute;
1.00: .../domain.te:154: allow untrusted_app
        security_file:dir { getattr search };
	\end{verbatim}
	\caption{An example of the \texttt{risky\_rules} plugin with the \textit{trust\_lh} scoring system.}
	\label{fig:risky-rules-plugin-trust-example}
\end{figure}

\subsubsection{Limitations}
Both scoring systems, \textit{risk} and \textit{trust}, assign a score to a rule by computing a formula over the partial scores of various rule elements.
These partial scores must be defined by an analyst in the plugin configuration file, and simply reflect what an analyst is most interested in.
Only the analyst who defined an element in the policy has the relevant knowledge to assign it a \textit{risk} or \textit{trust} score.
A high rule score does not mean that a rule is dangerous, and a low score does not mean that a rule is safe: a high score represents a rule which the analyst deems more interesting, and vice versa.

\subsection{Plugin 4: Unnecessary Rules}
\label{sec:unnecessary}
\textbf{Goal}
Some rules are effective only when used in combination.
For example, a \setype{type\_transition} rule is useless without the related \setype{allow} rules actually enabling the requested access.
Similarly, some permissions are meaningful only when granted in combination.
For example, an \setype{allow} rule which grants \setype{read} on a file type, without granting \setype{open} on the same type or \setype{use} on the related file descriptor type, will not actually allow the file to be read.
Another example is debug rules, which are effective only when used for an OEM internal engineering build, and should not be present in the derived user build which is actually shipped.
An analyst may want to check that all such rules are correctly wrapped inside debug M4 macros, which prevent them from appearing in the final user build.
The \texttt{unnecessary\_rules} plugin searches the policy for rules which are ineffective or unnecessary, as in the examples above.
It also looks for debug rules mistakenly visible in the user policy.

\noindent\textbf{Implementation}
The plugin provides 3 features: detection of ineffective rule combinations, detection of debug rules, and detection of ineffective permissions.

\noindent\emph{Ineffective rule combinations}:
The plugin detects missing rules from an ordered tuple of rules.
Tuples can be specified by an analyst in the plugin configuration file, and can contain placeholder arguments.
This functionality looks for rules matching the first rule in a tuple, and verifies that all other rules in the tuple are present in the policy.
An example is shown in Figure~\ref{fig:unnecessary-rules-plugin-1-example}.
The tuple contains three rules with placeholder arguments.
If a rule is found matching the first rule in the tuple, the arguments are extracted and substituted in the remaining rules; each of these rules must then be found in the policy.

\begin{figure}[t]
	\centering
	\small
	\begin{tabular}{l}
		\textbf{Tuple:}                                       \\
		\texttt{type\_transition \$ARG0 \$ARG1:file \$ARG2;}  \\
		\texttt{allow \$ARG0 \$ARG1:dir \{ search write \};}  \\
		\texttt{allow \$ARG0 \$ARG2:file \{ create write \};} \\
		\textbf{If found:}                                    \\
		\p\p\p\p\texttt{type\_transition a b:file c;}         \\
		\textbf{Look for:}                                    \\
		\p\p\p\p\texttt{allow a b:dir \{ search write \};}    \\
		\p\p\p\p\texttt{allow a c:file \{ create write \};}   \\
	\end{tabular}
	\caption{An example of the ``ineffective rule combinations'' functionality of the \texttt{unnecessary\_rules} plugin.}
	\label{fig:unnecessary-rules-plugin-1-example}
\end{figure}

\noindent\emph{Debug rules}:
The plugin detects rules containing debug types as either the domain or the type.
Debug types can be specified by an analyst in the plugin configuration file.

\noindent\emph{Ineffective permissions}:
The plugin detects rules which grant some particular permission on a type, but do not grant some other particular permission on that type or some additional permissions on some other (related) type.
All three sets of permissions can be specified by an analyst in the configuration file.
An example is shown in Figure~\ref{fig:unnecessary-rules-plugin-3-example}.
If any permissions from the first set are granted on a file, then either all the permissions in the second set must be granted on the file, or the permissions in the third set must be granted on the file descriptor.
The first rule grants \texttt{read} and \texttt{write} from the first set, and does not grant \texttt{open} from the second set; however, the second rule grants \texttt{use} on the file descriptor.
The constraint is therefore satisfied.

\begin{figure}[t]
	\small
	\begin{tabular}{l}
		\textbf{If found:}                                 \\
		\p\p\p\p\texttt{file \{ write read append ioctl\}} \\
		\textbf{Look for either:}                          \\
		\p\p\p\p\texttt{file \{ open \}}                   \\
		\textbf{or:}                                       \\
		\p\p\p\p\texttt{fd \{use\}}                        \\
		\textbf{Rules:}                                    \\
		\p\p\p\p\texttt{allow a b:file \{ read write \};}  \\
		\p\p\p\p\texttt{allow a b:fd use;}                 \\
	\end{tabular}
	\caption{An example of the ``ineffective permissions'' functionality of the \texttt{unnecessary\_rules} plugin.}
	\label{fig:unnecessary-rules-plugin-3-example}
\end{figure}

\noindent\textbf{Limitations}
The plugin allows an analyst to express very fine-grained information: this results in a somewhat complex configuration file.

\subsection{Plugin 5: User \texttt{neverallows}}
\textbf{Goal}
\setype{neverallow} rules can be used to specify permissions never to be granted in the policy.
For example, Google uses \setype{neverallow} rules extensively to prevent OEMs from circumventing core security structures of the policy.
However, \setype{neverallow} rules are only enforced at compile time in the normal \seandroid policy development workflow: this means that a policy change may be committed into an OEM's VCS, only to later find out that it infringes one or more \setype{neverallow} rules and therefore breaks the compilation.
The \texttt{user\_neverallows plugin} allows an analyst to define an additional set of \setype{neverallow} rules, and be able to check at any time if they are respected by the policy.
This can be very useful for OEM policy maintainers who would like to immediately make sure that developers contributing small policy changes do not introduce any undesired rules.
The plugin enforces a list of custom user-defined \setype{neverallow} rules on a policy, reporting any infringing rule.

\noindent\textbf{Implementation}
The plugin checks each rule in the policy which matches any user-specified \setype{neverallow}, and verifies that it does not grant any permission explictly forbidden in the \setype{neverallow}.
Custom \setype{neverallow} rules can be defined by the analyst in the plugin configuration file, in the same syntax as they would be written in the policy.

\noindent\textbf{Limitations}
The \setype{user\_neverallows} plugin processes each user-provided \setype{neverallow} rule individually: therefore, it works best with small numbers of rules (tens of thousands).

\section{\uppercase{Evaluation}}
\label{sec:evaluation}

\noindent In order to show that \selint fulfills the requirements stated in Section~\ref{sec:requirements}, we solicited feedback from \seandroid experts about their experience with \selint, as well as measured the tool's performance.

\subsection{Expert Survey}
\label{sec:userfeedback}

Following Requirement~\ref{req:config}, \selint is designed to be configured by an \seandroid expert before regular developers can use it in their work flow.
\seandroid experts are, therefore, the main target audience of \selint.
Developers are just expected to run \selint and verify that it doesn't produce new warnings on their policy modifications.
Thus, in order to evaluate the usability and usefulness of \selint, we need to collect feedback from \seandroid experts.

\noindent\textbf{Materials}
In order to collect expert feedback about \selint, we prepared an evaluation questionnaire\footnote{\ifanonymous Link anonymized for double-blind review \else \href{http://goo.gl/forms/j9oUBL2wnEjOvpLs2}{\texttt{goo.gl/forms/j9oUBL2wnEjOvpLs2}}\fi}.
\selint itself was available for download via our public Github repository\footnote{\ifanonymous Link anonymized for double-blind review \else \href{https://github.com/seandroid-analytics/selint}{\texttt{github.com/seandroid-analytics/selint}}\fi}.

\noindent\textbf{Procedure}
When collecting feedback on \selint, we wanted to focus on people that already have strong prior experience with \seandroid policies.
This choice is based on the fact that these experts are able to evaluate not just the tool itself, but also the default configuration we provide for its plugins.
In order to obtain such feedback, we announced the \selint tool on the \seandroid public mailing list\footnote{\href{https://www.mail-archive.com/seandroid-list@tycho.nsa.gov/}{\texttt{seandroid-list@tycho.nsa.gov}}}.
This mailing list is a common forum where discussions among \seandroid experts take place.
We asked people to fill in the questionnaire after trying to use the tool on their Android tree.

\noindent\textbf{Participants}
Three experts from three different companies evaluated \selint. Each had more than 2 years of experience with \seandroid policies.

\noindent\textbf{Results}
All respondents ranked \selint as easy to use, and its results as easy to interpret.
They also agreed that functionality offered by \selint is not currently provided by any existing tools; they ranked \selint as being ``\textit{valuable}'' for them for their current work on \seandroid.
Our free-form questions on the overall \selint experience gathered answers such as:
\begin{quote}
	\small
	``\textit{I was able to use the tool to find things I wanted to fix with respect to over-privileged domains and useless rules.}"
	
	``\textit{I think this just adds to the list of useful tools in policy development.
	The output is more user friendly than sepolicy-analyze and hopefully would appeal to those who only write policy infrequently - such as most OEMs.}"
\end{quote}

Out of all the default plugins we provided with \selint, the \setype{risky\_rules} plugin caught the most attention and received the most positive feedback.
This is as expected, given that this is the plugin that helps the most to directly evaluate the security of a \seandroid policy.
Plugins dealing with M4 macros were also found to be useful, with respondents reporting that they actually adopted most or all suggestions for \setype{global\_macros} or \setype{te\_macros} in their \seandroid policy.
The \setype{neverallow\_rules} plugin got an expected answer to the question ``\textit{Do you plan to use the neverallow\_rules plugin?}'':
\begin{quote}
	\small
	``\textit{Yes, to add rules I don't want in the policy, but where I don't want to add an actual neverallow.
		Neverallows end up in CTS, so you don't want to use them too much.
	As for OEM policy additions, sometimes neverallows are too strict and we just want to see what the linter picks up.}"
\end{quote}
This is exactly the usage we envisioned for it: an ability for OEMs to enforce custom \setype{neverallow} rules without them being checked by Android Compatibility Test Suite (CTS).
Respondents also had some good points for future enhancements, such as implementing an easier setup wizard and automatically prompting to input the scores for types or permissions which do not have one in the \texttt{risky\_rules} plugin.

\noindent\textbf{Limitations}
In order to perform a better evaluation of \selint, we need a more extensive study with many more OEM developers who need to modify \seandroid policies.
However, this is difficult to achieve because of the following reasons.
In order to try \selint, participants need to have their own custom Android tree and their own custom \seandroid policies, since the tool targets OEM \seandroid policy writers; this naturally limits the number of participants.
In addition, people that actually have their own custom policy are usually engineers working for OEMs.
They might not want to take part in our study because of corporate confidentiality concerns.
Another difficulty is in setting up \selint, as one of our respondents noted.
This is due to the fact that \selint relies on the policy representation library from SETools~\cite{setools4} to perform policy parsing, and older versions of this library do not support some new \seandroid policy elements, such as \setype{xperms}.
This, together with some compatibility issues between \seandroid policy versions and SETools, made it harder for some users to setup the tool initially.

Despite these limitations, we believe the user feedback we received confirms that our goals and assumptions for \selint and the default configurations of its plugins are correct.
In addition, this feedback gives us directions for future work discussed in Section~\ref{discussion}.
We also hope that we will receive more user feedback on our tool with time.

\subsection{Performance Evaluation}
\label{sec:performance}

\begin{table}[h]
	\caption{Performance measurements for \selint on Intel Android tree with 99532 expanded rules}
	\resizebox{\columnwidth}{!}{
		\begin{tabular}{lrr}
			\hline
			Component           & Avg time (s)      & Avg mem (MB)      \\
			\hline
			\selint core        & 0.40 $\pm$ 0.01   & 99.53 $\pm$ 0.06  \\
			user\_neverallows   & 0.43 $\pm$ 0.01   & 99.51 $\pm$ 0.05  \\
			simple macros       & 0.59 $\pm$ 0.02   & 99.94 $\pm$ 0.04  \\
			unnecessary\_rules  & 0.65 $\pm$ 0.01   & 99.52 $\pm$ 0.08  \\
			risky\_rules        & 1.06 $\pm$ 0.01   & 99.51 $\pm$ 0.05  \\
			parametrized macros & 168.42 $\pm$ 2.17 & 446.52 $\pm$ 0.07 \\
			\hline
		\end{tabular}
	}
	\label{tab:performance-intel}
\end{table}

\begin{table}[h]
	\caption{Performance measurements for \selint on \aosp tree with 3081233 expanded rules}
	\resizebox{\columnwidth}{!}{
		\begin{tabular}{lrr}
			\hline
			Component           & Avg time (s)        & Avg mem (MB)       \\
			\hline
			\selint core        & 1.88 $\pm$ \p0.02   & 212.11 $\pm$ 0.07  \\
			user\_neverallows   & 1.89 $\pm$ \p0.02   & 212.09 $\pm$ 0.07  \\
			simple macros       & 2.18 $\pm$ \p0.03   & 219.03 $\pm$ 0.09  \\
			unnecessary\_rules  & 20.25 $\pm$ \p0.17  & 212.07 $\pm$ 0.06  \\
			risky\_rules        & 3.23 $\pm$ \p0.03   & 212.07 $\pm$ 0.07  \\
			parametrized macros & 3210.03 $\pm$ 48.13 & 6031.84 $\pm$ 0.59 \\
			\hline
		\end{tabular}
	}
	\label{tab:performance-aosp}
\end{table}

In order to evaluate the performance of \selint we conducted a set of measurements, collecting execution time and memory usage.
We consider these numbers to be the most important indicators for \selint, since it can be used either manually by a single person or automatically as part of a Continuous Integration (CI) process.
The measurements were conducted on an off-the-shelf laptop with an Intel Core i7-4770HQ 2.20GHz CPU and 16GB of 1600MHz DDR3 RAM.
Each measurement was repeated 10 times, and the average and standard deviation are presented in Table~\ref{tab:performance-intel} and Table~\ref{tab:performance-aosp}.
The first table presents data for a public Intel tree, Android 5.1\footnote{\href{https://github.com/android-ia}{\texttt{github.com/android-ia}}}, and the second one for the public \aosp tree, master branch\footnote{\href{https://android.googlesource.com}{\texttt{android.googlesource.com}}}.
For all measurements we have measured the \selint core and each of its plugins separately.
The big difference in performance between these two trees comes from the number of expanded rules in the source policies: for the Intel tree it is 99532, while for the \aosp tree it is 3081233.
The execution time of the \texttt{unnecessary\_rules} plugin scales differently than others, requiring almost the same time as the \selint core on the Intel tree and 10 times more than the \selint core on the \aosp tree.
This is due not only to the different total number of rules in the two trees, but also to the number of rules that each domain has, since the plugin needs to check for ineffective rule combinations or permissions (see Section~\ref{sec:unnecessary}).
The \texttt{parametrized macro} plugin is the only plugin that takes a considerable amount of time to run, especially on the \aosp tree.
As explained in Section~\ref{sec:te-macro}, this is due to the fact that we are currently not implementing any heuristics in our solution to the problem, and are just relying on exploration of the solution space.
As a result, the current plugin should not be included into the default set of plugins executing automatically as part of a CI process, but should be used manually by an expert.
The execution time and memory usage of the other plugins fit the desired use cases: given that normally an \aosp build takes at least half an hour to complete in a powerful CI infrastructure, an overhead of minutes and hundreds of MB of memory is considered acceptable.

\section{\uppercase{Discussion}}
\label{discussion}
\noindent While our evaluation showed that \selint is considered a valuable tool for analyzing \seandroid policies, there are many areas for future work and improvements.
The initial setup of \selint would benefit from an interactive procedure, allowing users to automatically detect and solve the possible mismatches between the installed libraries and policy versions.
The \texttt{parametrized macro} plugin could provide an implementation based on a heuristic solution for the knapsack problem allowing users to obtain a partial solution, in order to save time and enable this plugin to be run as part of a CI infrastructure.
More work is needed in order to polish the default configuration offered by the \texttt{risky\_rules} plugin, and to provide a way for OEMs to easily, and maybe interactively, add scores for their own domains and types.
We also need to conduct a study on how easy it is for \seandroid experts to write new \selint plugins.
Another future research direction is to investigate the possibility of using \selint together with a policy decompiler, in order to analyze OEM policies from available Android devices.
This would provide additional input for \selint evaluation.

We continue to gather feedback from \selint users and \seandroid experts to adjust \selint to their needs and requirements.
Since \selint is open source software, and builds on existing official \seandroid tools, we are planning to work with Google to include \selint in the set of \seandroid tools provided with the \aosp tree.

%ACKNOWLEDGMENTS are optional
%\section{Acknowledgments}
%The authors would like to thank William Roberts, for his contributions to \selint design and evaluation; Mika Juuti, for his helpful discussions about the knapsack problem; all the users who tried \selint and filled our evaluation questionnaire, for their feedback and suggestions.
%This work is partially supported by the Intel Collaborative Research Institute for Secure Computing (ICRI-SC)\footnote{\href{http://www.icri-sc.org/}{\texttt{icri-sc.org}}}.

\bibliographystyle{apalike}
{\small
	\bibliography{paper}}

\end{document}